\documentstyle [12pt] {article}
\title {A "QUASI-RAPID" EXTINCTION POPULATION DYNAMICS AND MAMMOTHS OVERKILL}

\author {Vladan Pankovi\'c $^{\ast,\sharp}$, Rade Glavatovi\'c $^{\Diamond}$, Nikola Vunduk
$^{\Box}$,\\ Dejan Banjac$^{\sharp}$, Nemanja Marjanovi\'c
$^{\sharp}$,
Milan Predojevi\'c $^{\ast,\sharp}$\\
$^{\ast}$ Department of Physics, Faculty of Sciences \\21000 Novi Sad, Trg Dositeja Obradovi\'ca 4,  Serbia \\
$^{\sharp}$ Gimnazija, 22320 Indjija, Trg Slobode 2a \\ Serbia, vladanp@gimnazija-indjija.edu.yu\\
$^{\Diamond}$Military Medical Academy, 11000 Beograd, Crnotravska 17, Serbia\\
$^{\Box}$Osnovna \v{s}kola "Petar Ko\v{c}i\'c", 22320 Indjija,
\\Cara Du\v{s}ana 4, Serbia, nikola.v@neobee.net}

\date {}
\begin {document}
\maketitle

\vspace {1.5cm}

\begin {abstract}

In this work we suggest and consider an original, simple
mathematical model of a "quasi-rapid" extinction population
dynamics. It describes  a decrease  and final extinction  of the
population of one prey species by a "quasi-rapid" interaction with
one predator species with increasing population. This
"quasi-rapid" interaction means ecologically that prey species
behaves practically quite passively (since there is no time for
any reaction, i.e. defense), like an appropriate environment, in
respect to "quasi-rapid" activity of the predator species that can
have different "quasi-rapid" hunting  abilities. Mathematically,
our model is based on a non-Lotka-Volterraian system of two
differential equations of the first order, first of which is
linear while second, depending of a parameter that characterizes
hunting ability is nonlinear. We compare suggested "quasi-rapid"
extinction population dynamics and the global model of the
overkill of the prehistoric megafauna (mammoths). We demonstrate
that our "quasi-rapid" extinction population dynamics is able to
restitute successfully correlations between empirical
(archeological) data and overkill theory in North America as well
as  Australia. For this reason, we conclude that global overkill
theory, completely mathematically modelable by "quasi-rapid"
extinction population dynamics can consistently explain the
Pleistocene extinctions of the megafauna.
\end {abstract}
\vspace {1.5cm}

\section {Introduction}

As it is well-known [1]-[3], Pleistocene extinctions of the
megafauna (mammoths, etc.) represent a significant, open problem
in the paleontology. There are two main mutually opposite attempts
of the  solution of given problem. First one, so-called (global)
overkill (blitzkrieg) theory [4]-[6], supposes that human
activity, precisely a rapid hunting or overhunting of  the
megafauna realized by initially relatively small but relatively
quickly movable human groups - immigrants, has had dominant role
in these extinctions. Second one, so-called climate changes theory
[7]-[10], suggests that climate changes have had dominant role in
the extinctions of the Pleistocene megafauna. Overkill theory
points out that megafauna extinctions occurred in the different
areas in the different times which, from overkill theory view
point, eliminates climate changes as primary factors of these
extinctions. Also, overkill theory point out that any megafauna
extinction started "immediately" after arrival of the human
hunting groups in given area (eg. Clovis people immigration in
North America) and correlated with relatively rapid migration
(blitzkrieg or a wave like front) of given group over given area.
All this in many cases corresponds to  archeological data. Climate
changes theory pointed out that in some cases, eg. in Australia,
according to most recent archeological data [11], there is a
relatively long-lasting period of the coexistence of the human
hunting groups and Pleistocene megafauna. From the climate changes
theory view point, it eliminates human hunting as primary factor
of the megafauna extinctions. Also, there are observations [12],
[13] that extinction interaction between human hunting groups as
predators and megafauna as preys, supposed within overkill theory,
contradicts to basic  population dynamics, i.e. Lotka-Volterra
equations [14]-[18] according to which both, predator species
population and prey species population, oscillate but not
disappear during time. Even if there are different computer
simulations of the overkill theory in North America, eg. [19],
they cannot remove given contradiction as well as they cannot
explain Pleistocene megafauna extinction in Australia. Finally,
there are different attempts of the explanation of the Pleistocene
megafauna extinctions  that suppose that these extinctions have
been caused by a combination of the human hunting and climate
changes  [20], [21] or by hyperdisease [22], or by a very complex
population dynamics based on the many species Lotka-Volterra
equations [13], etc. Meanwhile, it seems that none of the
mentioned attempts and theories is completely successful.

In this work we shall suggest and consider an original, simple
mathematical model of an extinction population dynamics. It
describes  a decrease  and final extinction  of the population of
one prey species by a "quasi-rapid" interaction with one predator
species with increasing population. This "quasi-rapid" interaction
means ecologically that prey species behaves, practically, quite
naively, precisely, passively or quasi-passively (since there is
no sufficient time for any reaction, "response", i.e. defense
tactic), like an appropriate environment, in respect to
"quasi-rapid" activity of the predator species. Also, different
predator species can have different "quasi-rapid" hunting
abilities. Mathematically, our model is based on a
non-Lotka-Volterraian system of two differential equations of the
first order. This principal distinction between our and
Lotka-Volterra equation system is ecologically quite reasonable.
Namely,  Lotka-Volterra equations system refers on a "slow"
interaction between one predator species and one prey species.
This "slow" interaction means ecologically that prey species, in
respect to "slow" predator species activity, behaves nonnaively,
precisely, actively  (since there is sufficiently time for a
significant  reaction or "response", i.e. defense tactic modelable
by corresponding Lotka-Volterra cross terms), unlike a naive and
passive  environment. Our first equation is linear while second,
depending of a hunting ability parameter that, simply speaking,
characterizes hunting ability, is nonlinear. (But in the limit
when hunting ability parameter tends to its maximal value 1, the
second equation and whole equations system become linear too.
Such  linearity induces a population superposition principle. It
means mathematically that a sum of two populations of one species,
any of which satisfies given equations system, satisfies given
equations system too. It characterizes a wave like population
change. Our equations system describes in simple manner the
populations homogeneously distributed over surface of given area,
i.e. populations that do not depend of the space coordinates. But
even in this case mentioned population superposition principle can
implicitly support a wave like front of the migrations of the
human hunting groups over surface of given area.)  Since, in this
way, whole our equations system depends, in fact, of the hunting
ability parameter that can have different values, it can be
concluded that this system can be applied in many different
ecological situations any of which represents an especial case of
the "quasi-rapid" interaction between one predator species and one
prey species.We shall compare suggested "quasi-rapid" extinction
population dynamics and the global model of the overkill of the
Pleistocene megafauna (mammoths). We shall demonstrate that our
"quasi-rapid" extinction population dynamics is able to restitute
successfully correlations between empirical (archaeological) data,
i.e. estimated parameters and predictions or intentions of the
global overkill theory, especially in North America and Australia.
For this reason, we shall conclude that global overkill theory,
completely mathematically modelable by "quasi-rapid" extinction
population dynamics, can consistently explain the extinction of
the Pleistocene megafauna. In other word, instead of a "mixture"
of the overkill and climate change influences at the extinction of
the Pleistocene megafauna, a "mixture" of the "quasi-rapid"
extinction population dynamics for different values of the hunting
ability parameter is completely sufficient for consistent
explanation of the extinctions of the Pleistocene megafauna in
full agreement with global overkill theory.

\section {A "quasi-rapid" extinction population dynamics}

We shall suggest the following system of the differential
equations
\begin {equation}
\frac {dx}{dt} = - {\it a} \frac {dy}{dt}
\end {equation}
\begin {equation}
\frac {dy}{dt} = {\it b}y^{k}  \hspace{1cm}    for \hspace{0.5cm}
0 < k \leq 1
\end {equation}
where $x$, $y$ represent the real positive variables that depend
of the time $t$, while ${\it a}$, ${\it b}$ and ${\it k}$
represent the time independent real, positive constants, i.e.
parameters.

We shall suppose that given system can be used for mathematical
modeling of a population dynamics. Namely, we shall suppose that
$x$ represents the population of a prey species and that y
represents the population of a predator species.

Further, it can be observed that for $k$ equivalent to 1, i.e. for
\begin {equation}
   k = 1
\end {equation}
(2) turns in the following linear differential equation
\begin {equation}
   \frac {dy}{dt} = {\it b}y
\end {equation}
with simple solution
\begin {equation}
    y = y_{0}\exp({\it b}t)
\end {equation}
where  $y_{0}$  represents the initial value of $y$. It, in fact ,
corresponds to  well-known [14]-[18] unlimited increase of the
population of a species placed in an appropriate environment (that
behaves quite naively or passively, without any reaction or
"response", i.e. defense tactic, in respect to given species
activity). Then  ${\it b}$ corresponds to so-called  birth rate of
given species that represents simply the interaction between given
(predator) species and environment. It suggests that prey species
must implicitly correspond to given environment.

For this reason we shall suppose that even for $k$ smaller than 1
but relatively close to  1, i.e. for
\begin {equation}
    k \leq 1
\end {equation}
nonlinear differential equation (2), precisely its left hand,
represents ecologically  the "speed" of the predator species
population corresponding to a species placed in an appropriate
environment corresponding implicitly to prey species. Right hand
of (2) points out that population "speed" is equivalent to a power
function of the population so that this "speed" is larger and
larger for $k$ closer and closer to 1. In this way  $k$ can be
considered as a degree of the population "speed" or degree of the
"quasi-rapid" interaction between predator species and
environment, or, implicitly, prey species. Also, it can be
considered that $k$ expresses implicitly a hunting ability of the
predator species. (Detailed ecological analysis of the  hunting
ability parameter goes over basic intentions of this work.
Intuitively, it would be expected that  this parameter depends not
only of the characteristics of the predators, eg human, species,
but also of the geographical characteristics, eg. magnitude of the
surface of given Earth area, etc. It would be suspected that
hunting ability parameter increases when surface of given are
decreases which can explain extremely rapid extinction of a prey
species at small islands.)  Especially, in the  limit (2), given
interaction can be called "rapid". In other words we shall suppose
that when interaction between one prey species and one predator
species is "quasi-rapid" (which means that prey species behave,
practically, quite passively or quasi-passively,without any
reaction, "response", i.e. defense tactic, in respect to action of
the predator species ) predator species population dynamics can be
presented by (2)   for $k$ relatively close to 1. Simple solution
of (2) in this case is
\begin {equation}
   y = (y_{0}^{1-k} + (1-k){\it b}t)^{\frac {1}{1-k}}             .
\end {equation}

It can be observed and pointed out that (2) and (7) depend
principally of the hunting ability parameter so that for different
values of this parameter, any of which is relatively close to 1,
there are different but "quasi-rapid" increase of the predator
species population.

Equation (1) simply means that "speed" of the prey species
population is proportional to negative "speed" of the predator
species population. The same equation, independently of the value
of $k$, can be simply transformed in
\begin {equation}
   \frac {d(x +{\it a}y)}{dt} = 0
\end {equation}
which yields
\begin {equation}
   x + {\it a}y = x_{0} + {\it a}y_{0} = const
\end {equation}
where  $x_{0}$ represents initial prey species population. In
other words expression $x+{\it a}y$ represents a form that stands
conserved during time  so that (9) can be considered as a
conservation law. Namely,  ${\it a}y$ can be considered as the
{\it  calibrated (dilated)}  predator species population  and, in
this sense, (9) can be considered as the {\it law of the
population conservation} (prey species population turns in the
predator species calibrated population, but whole population
representing sum of the prey species population and predator
species calibrated population stands conserved during time). For
the same reason ${\it a}$ can be called calibration parameter.

From (9) it follows
\begin {equation}
   x =  x_{0} + {\it a}y_{0} - {\it a}y
\end {equation}
which, for (3),(5) yields
\begin {equation}
   x = x_{0} + {\it a}y_{0}  - {\it a}y_{0} \exp({\it b}t)
\end {equation}
and, for (6),(7),
\begin {equation}
   x = x_{0} + {\it a}y_{0} - {\it a}(y_{0}^{1-k}+(1-k){\it b}t)^{\frac {1}{1-k}}
\end {equation}

It is not hard to see that predator species population  (5) or (7)
represents a monotonously increasing time function while prey
species population (11) or (12) represents a monotonously
decreasing time function. It implies that there is a finite time
moment, called extinction time, $T$, in which prey species
population becomes equivalent to zero, i.e.
\begin {equation}
   x(T) = 0
\end {equation}
For (11) it yields
\begin {equation}
   T = \frac {1}{b} \ln (\frac {x_{0}}{{\it a}y_{0}} + 1)
\end {equation}
while for (12) it yields
\begin {equation}
  T = \frac {1}{{\it b}(1-k)}
((\frac {x_{0}}{{\it a}} + y_{0})^{1-k} - y_{0}^{1-k})
\end {equation}

In this way we obtain an original, simple mathematical model of a
population dynamics with decrease and final extinction of the
population of one prey species in a finite time moment by
"quasi-rapid" interaction with one predator species with
increasing population. This model is principally different from
the usual Lotka-Volterra equations system of two nonlinear
differential equations that describes "slow" interaction between
one predator and one prey species. Namely,  "slow" interaction
means ecologically that prey species, in respect to "slow"
predator species activity, behaves nonpassively (nonnaively),
i.e.  actively  (since there is sufficiently time for a
significant  reaction or "response", i.e. defense tactic modelable
by corresponding Lotka-Volterra cross terms), unlike a passive
environment. It, on the one hand, implies that hunting ability
parameter $k$ in (2) becomes significantly smaller than 1, and, on
the other hand, that system (1),(2) corresponding to "quasi-rapid"
extinction population dynamics {\it cannot be applied at all}  for
description of the "slow" interaction between predator species and
prey species. (In this work we shall not analyze it with more
details, from ecological view point, what is meaning of the
expression that $k$ is relatively close to 1 or that $k$ is
significantly smaller than 1. We shall use the following rough or
ad hoc criterion: $k$ is relatively close to 1 for $k > 0.5$ and
vice versa $k$ is significantly smaller than 1 for $k \leq 0.5$. )
Since our equations system, in fact, depends of the hunting
ability parameter, as well as birth rate parameter and calibration
parameter, that can have different values, it can be concluded
that this system can be applied in many different ecological
situations any of which represents an especial case of the
"quasi-rapid" interaction between one predator species and one
prey species.  Especially, it can be observed that our equations
system (1), (2), i.e. its solutions (7), (12) have the following
important characteristics. They are more sensitive in respect to
variation of the hunting ability parameter than variations of
other parameters and initial populations.  Or, relative large
variations of the other parameters and initial populations
(corresponding to relatively large uncertainties of corresponding
empirical, i.e. archeological data and estimations) can be
relatively simply compensated by relatively small variations of
the hunting ability parameters. In other words,  many relatively
large uncertainties of the empirical (archeological) data are
practically irrelevant for distinction between a "quasi-rapid" and
"slow" interaction between one predator species and one prey
species. All this opens a possibility that different Pleistocene
megafauna extinctions, eg.  Pleistocene mammoths extinction in
North America and megafauna extinction in Australia, would be
consistently mathematically modeled by "quasi-rapid" extinction
population dynamics for different values of the hunting ability
parameters (any of which is relatively close to 1). If given
possibility would be affirmed then it can be concluded that basic
suppositions,  predictions and intentions of the global overkill
theory, completely mathematically modelable by "quasi-rapid"
extinction population dynamics, can consistently explain the
Pleistocene extinctions of the megafauna. In other words, it would
mean, according to basic supposition of the global overkill
theory, that  human hunting activities played dominant role while
climate changes have had only secondary role in the Pleistocene
extinctions of the megafauna.

\section {"Quasi-rapid" extinction population dynamics \\ and Pleistocene mammoths overkill
 in North America}

Now we shall attempt to apply "quasi-rapid" extinction population
dynamics (1),(2) for mathematical modeling of the global overkill
theory  [4] - [6]. Firstly, we shall attempt to apply
"quasi-rapid" extinction population dynamics (1), (2) for
mathematical modeling of the hypothesis on the Pleistocene
mammoths overkill (blitzkrieg) in North America [1], [2], [4]-[6],
[19], [23]. In other words we shall attempt to restitute, by
equations system (1), (2) consistent correlations between
empirical (archeological) data and estimated parameters that
characterize Pleistocene extinction of the megafauna in North
America.

So, suppose  that Clovis population increased about $3\%$ for one
year. It, introduced in (5), yields
\begin {equation}
  {\it b} = \ln (1.03)  (yr^{-1}) = 0.0295 (yr^{-1})
\end {equation}

Suppose, further, that a typical small group of about
\begin {equation}
  y_{0} = 50
\end {equation}
Clovis people killed about 15 mammoths per year. It also means
that a group of $x_{0}$ mammoths contacted during one year by
given small group of Clovis people was reduced in the group of
$x_{0} - 15$ mammoths. Introduction of this supposition and (16)
in (11), which implies "rapid" extinction population dynamics,
yields
\begin {equation}
   x_{0} - 15 = x_{0} + 50 {\it a}  - 50 {\it a} \cdot 1.03 = x_{0}  -  1.5 {\it a}
\end {equation}
or
\begin {equation}
    {\it a} = \frac {15}{1.5} = 10
\end {equation}

Suppose, finally,  that mammoth extinction occurred in the
extinction time interval that equals about 400 years, i.e.
\begin {equation}
   T = 400 (yr)
\end {equation}
 It, introduced, in common with (16) and (19), in (15) yields
\begin {equation}
    400 = \frac {1}{\ln (1.03)} \ln (\frac {1}{10}\frac {x_{0}}{y_{0}}+ 1)
\end {equation}
or
\begin {equation}
  \frac {x_{0}}{y_{0}} = 10 \exp (400 \ln (1.03)) - 1 = 1.36 \cdot 10^{6}
\end {equation}
where $x_{0}$ represents the initial population of the mammoths
while  $y_{0}$ represents the initial population of the Clovis
people. From (22) it follows
\begin {equation}
  x_{0} =  1.36 \cdot 10^{6} y_{0}
\end {equation}
which, for supposed initial Clovis population (17), yields
\begin {equation}
  x_{0} =  68 \cdot 10^{6}
\end {equation}
It represents a number comparable, precisely  about 10 times
greater than  roughly estimated initial mammoths population before
appearance of the Clovis people
\begin {equation}
  x_{0} =  10 \cdot 10^{6}  = 10^{7}                                   .
\end {equation}
In other words, "rapid" extinction population dynamics (1), (2)
very roughly correlates the existing empirical (archeological)
data and suppositions of the overkill theory in the case of the
Pleistocene mammoths extinction in North America.

Suppose, meanwhile, that Pleistocene mammoths overkill in North
America can be more successfully modeled by "quasi-rapid"
extinction population dynamics.

Suppose, also,
\begin {equation}
  x_{0} = 10^{7}
\end {equation}
\begin {equation}
  y_{0} = 10^{2}
\end {equation}
\begin {equation}
  {\it a} = 10
\end {equation}
\begin {equation}
  {\it b} = 2.5 \cdot 10^{-2}
\end {equation}
\begin {equation}
   T = 400 (yr) = 4 \cdot 10^{2} (yr)
\end {equation}
comparable with  (25), (17), (19), (16), (20). Introduction of
(26)-(30) in (15) yields
\begin {equation}
  (1-k) \simeq \frac {1}{10}(10^{6(1-k)}   -  10^{2(1-k)} )
\end {equation}
(First term on the right hand of (31) is obtained  by neglecting
of term $y_{0}=10^{2}$ relatively small in respect to term
$x_{0}/{\it a} = 10^{6}$). It represents a transcendent algebraic
equation whose solution can be obtained in the following way.
First of all it is well-known the following
\begin {equation}
   10 = \exp ( \ln (10)) \simeq \exp (2.3)
\end {equation}
which, introduced in (31), yields
\begin {equation}
       (1-k) \simeq \frac {\exp ( 13.8 (1-k))}{10^{1}} - \frac { \exp {4.6 (1-k)}}{10^{1}}.
\end {equation}
Further, for "quasi-rapid" interaction, according to its
definition, $k$ is close to 1 and $1-k$ to 0. It implies that both
hands of (33) are close to 0, while both exponential terms at
right hand of (33) are close to 1. It admits that given terms can
be Taylor expanded in the quadratic approximation which introduced
in (33) yields
\begin {equation}
  (1-k) \simeq  0.92 (1-k) + 16.92 (1-k)^{2}     .
\end {equation}
It represents an algebraic quadratic equation whose unique
solution , since $k$ must be close to 1 and $1-k$  must be
positive, is
\begin {equation}
  k \simeq 0.995
\end {equation}
\begin {equation}
  1-k \simeq 0.005 .
\end {equation}

So, "quasi-rapid" extinction population dynamics (1),(2), for $k$
(35) really close to 1, can successfully correlate all estimated
significant data (26)-(30) that characterize Pleistocene mammoths
extinction in North America according to overkill (blitzkrieg)
theory. Also, it is not hard to see, according to previously
mentioned characteristic of the equations system (1),(2), that
possible relatively large uncertainties and variations of the
empirical (archeological) data and estimated parameters (26)-(30)
, are practically irrelevant for final conclusion that $k$ must be
close to 1.

\section {A "quasi-rapid" extinction population dynamics\\ and overkill  of the Pleistocene mega-fauna in Australia}

Now, we shall attempt to apply "quasi-rapid" extinction population
dynamics (1),(2) for mathematical modeling of the hypothesis on
the Pleistocene megafauna extinction in Australia. In other words
we shall attempt to restitute, by equations system (1),(2),
consistent correlations between empirical (archeological) data and
estimated parameters that characterize Pleistocene extinction of
the megafauna in Australia. According to recent Roberts et {\it
al.} data [11] human population arrived in Australia before $56
000 \pm 4 000$ yr, while extinction of the Pleistocene megafauna
in Australia occurred before $46 000 \pm 5000$ yr. It implies that
Pleistocene megafauna extinction in the Australia  occurred in an
uncertainly determined  time interval, i.e. extinction time $T$
that equals about $10 000 \pm 9 000$ yr. Suppose, meanwhile, that
real
   value of $T$ is very close to its mean value $10 000$ yr, i.e..
\begin {equation}
   T = 10 000  (yr)              .
\end {equation}

Suppose that given megafauna extinction can be mathematically
modeled by "quasi-rapid" extinction population dynamics (1),(2).

Suppose, also,
\begin {equation}
  x_{0} = 10^{7}
\end {equation}
\begin {equation}
  y_{0} = 10^{2}
\end {equation}
\begin {equation}
  {\it a} = 10
\end {equation}
\begin {equation}
  {\it b} = 2.5 \cdot 10^{2}
\end {equation}

Obviously, (38)-(41) are equivalent to (26)-(29). It represents a
reasonable supposition. But, extinction time (37) is significantly
larger than extinction time (30), which implies that here $k$
value must be significantly different from 0.995 (35).

Introduction of (37)-(41) in (15) yields
\begin {equation}
  (1-k) 250 = 10^{6(1-k)}  - 10^{2(1-k)}
\end {equation}
It represents a transcendent equation. For reason of relatively
large value of T (37), we shall solve (42) simply numerically, by
fitting, which, with accuracy of  $0.1\%$  yields
\begin {equation}
    k = 0.680  > 0.5
\end {equation}
It can be considered as a value relatively close to 1 (in sense
that it is greater than 0.5) so that supposition on the
applicability of the "quasi-rapid" population dynamics (1),(2) can
be considered consistent. On the other hand this value is
relatively small which causes a relatively small "sped" of the
megafauna population and relatively large extinction time (37).

So, "quasi-rapid" extinction population dynamics (1),(2), for $k$
(42) , can successfully to correlate all estimated significant
data (26)-(30) that characterize Pleistocene extinction of the
megafauna in Australia according to general (global) overkill (but
not blitzkrieg) theory. Again it is not hard to see that possible
uncertainties and variations of the empirical (archeological) data
and estimated parameters (37)-(41), are practically irrelevant for
final conclusion that $k$ must be relatively close to 1.

\section {Conclusion}

 In conclusion we can shortly repeat and point out the following. In this work we suggested and considered an original, simple mathematical model
 of a "quasi-rapid" extinction population dynamics. It describes  a decrease  and final extinction  of the population of one prey species
 by a "quasi-rapid" interaction with one predator species with increasing population. This "quasi-rapid" interaction means ecologically
 that prey species behave, practically, quite passively (since there is no sufficient time for a significant  reaction or "response",
 i.e. defense tactic), like an appropriate environment, in respect to activity of the predator species that can have different "quasi-rapid"
 hunting  abilities. Mathematically, our model is based on a non-Lotka-Volterraian system of two differential equations of the first order,
 first of which is linear while second, depending of a parameter that characterizes hunting ability, is nonlinear .
 Global overkill scenario can be mathematically completely modeled and in this sense affirmed by suggested  "quasi-rapid" extinction
 population dynamics  even if in different cases (overkill in North America, Australia, etc.) corresponding hunting ability
 parameter can have different values (smaller than 1, but relatively close to 1). It implies that human hunting activities
 played dominant role while climatic changes have had only secondary role in the extinctions of the Pleistocene megafauna.
 In other word, instead of a "mixture" of the overkill and climate change influences, a "mixture" of the "quasi-rapid"
 extinction population dynamics for different values of the hunting ability parameter is completely sufficient for consistent
 explanation of the Pleistocene extinctions of the megafauna in full agreement with global overkill theory.

\newpage

\section {References}

\begin {itemize}
\item [ [1] ]  {\it Pleistocene Extinctions : The Search of a Cause }, eds. P.S.Martin, H.E.Wright (Yale University, New Havan, 1967.)
\item [ [2] ] {\it Quaternary Exstinctions : A Prehistoric Revolution }, eds. P.S.Martin, G.Klein (University of Arizona Press, Tucson, 1984.)
\item [ [3] ]  {\it Extinctions in Near Time : Causes, Context and Consequences }, ed. R.D.E. MacPhee (Kluewer Academic-Plenum Press, New York, 1999.)
\item [ [4] ]  P.S.Martin, {\it Prehistoric Overkill }, in [1], 75.
\item [ [5] ]  J.Moismann, P.S.Martin, Am.Sci., {\bf 63}, (1975.), 304.
\item [ [6] ]  P.S.Martin, {\it Prehistoric Overkill : The Global Model}, in [2], 354.
\item [ [7] ]  R.D.Guthrie, {\it Mosaics, Allochemics and Nutrients :  an Ecological Theory of Late Pleistocene Megafaunal Extinctions }, in [2], 259.
\item [ [8] ]  R.D.Guthrie, {\it Frozen Fauna of the Mamoth Stepe : The Story of Blue Babe } (Chicago University Press, Chicago, 1990.)
\item [ [9] ]  R.W.Graham, E.L.Lundelius, {\it Coevolutionary Disequilibrium and Pleistocene Extinctions }, in [2], 223.
\item [ [10] ]  D.K. Grayson, {\it Late Pleistocene Mammalian Extinction in North America : Taxonomy, Chronology, and Explanations },
Journal of World Prehistory, {\bf 5}, (1991.), 193.
\item [ [11] ]  R.G.Roberts,  T.F.Flannery, L.K.Ayliffe,  H.Yoshida, J.M.Olley, G.J.Predaux, G.M.Laslett, A.Baynes, M.A.Smith, R.Jones, B.L.Smit, Science, {\bf 292}, (2001.), 1888.
\item [ [12] ]  G.E.Belovsky, J.Antrophol.Archeol., {\bf 7}, (1988.), 329.
\item [ [13] ]  E.Whitney-Smith, {\it Second Order Predation and Pleistocene Extinctions : A System Dynamics Model }, Ph.D.,
http://quaternary.net/extinct2000/
\item [ [14] ]  R.M.May, {\it Stability and Competition in Model Ecosystems } (Princeton Univ.Press., Princeton, New Jersey,1974.)
\item [ [15] ]  E.C.Pielou, {\it Mathematical Ecology } (John Wiley and Sons, New York, 1977.)
\item [ [16] ]  F.Verhulst, {\it Nonlinear Differential Equations and Dynamical Systems }(Springer Verlag, Berlin, 1990.)
\item [ [17] ]  J.D.Murray, {\it Mathematical Biology } (Springer Verlag, Berlin-Heidelberg, 1993.)
\item [ [18] ]  D.Alstad, {\it Basic Populs Models in Ecology} (Prentice-Hall Inc., New York, 2001.)
\item [ [19] ]  J.Alroy, {\it A Multispecies Overkill Simulation of the End-Pleistocene Magafaunal Mass Extinction }, Science, {\bf 292}, (2001.), 1893.
\item [ [20] ]  A.J.Stuart, {\it Mammalian Extinction in the Late Pleistocene of Northern Euroasia and North America }, Biological Reviews, {\bf 66}, (1991.), 453.
\item [ [21] ]  A.J.Stuart, {\it Late Pleistocene Megafaunal Extinctions ; a European Perspective }, in [3].
\item [ [22] ]  R.D.E MacPhee, P.A.Marx, {\it The 40 000 Year Plague ; Humans, Hyperdisease and First Contact Extinction },
in, {\it Natural Change and Human Impact in Madagascar}, eds. S.M.Goodman, B.D.Patterson (Smithsonian Institute Press, Washington D.C., 1997.), 169.
\item [ [23] ]  J.Steele, J.Adams, T.Sluckin,World Archeol., {\it 30}, (1998.), 286.

\end {itemize}

\end {document}